\documentstyle[12pt]{article}

\begin{document}
\input epsf.tex
\rightline{McGill/96-12}
\vspace{.5cm}
\begin{center}
{\Large\bf A Multiple Commutator Formula for the Sum of Feynman Diagrams}\\
\bigskip\bigskip\bigskip
{C.S. Lam$^*$ and K.F. Liu$^\dagger$}\\
\bigskip
{\it $^*$Department of Physics, McGill University,\\
3600 University St., Montreal, P.Q., Canada H3A 2T8\\
Email: Lam@physics.mcgill.ca.\\}
{\it and\\}
{\it $^\dagger$Department of Physics and Astronomy,
University of Kentucky\\
Lexington, KY 40506, U.S.A.\\
Email: Liu@ukcc.uky.edu .}
\end{center}
\begin{abstract}


In the presence of a large parameter, such as mass or energy,
leading behavior of individual
Feynman diagrams often get cancelled in the sum. This is
known to happen in large-$N_c$ QCD in the presence of a baryon, and
also in the case of high-energy electron-electron as well as quark-quark
scatterings. We present an exact combinatorial formula, involving
multiple commutators of the vertices, which can be used to compute
such cancellations. It is a non-abelian generalization
of the eikonal formula, and will be applied in subsequent publications
to study the consistency of large-$N_c$ QCD involving baryons, as well
as high-energy quark-quark scattering in ordinary QCD.
\end{abstract}

\section{Introduction}

This work arose from the challenge to resolve a possible inconsistency
of large-$N_c$ QCD in the presence of baryons \cite{GH,JL,CC,MP,YG}.
Briefly the problem is as follows. The meson-baryon Yukawa
coupling constant is proportional to $\sqrt{N_c}$, so an $n$-meson baryon
tree diagram grows as $N_c^{n/2}$. On the other hand,
the complete $n$-meson baryon amplitude must decrease with
$N_c$ like $N_c^{1-n/2}$ \cite{GH}. This means that an individual Feynman
diagram is
a factor $N_c^{n-1}$ larger than the sum of the $n!$ permuted diagrams! To be
consistent, a huge cancellation must take place in the sum of diagrams, and
the larger $n$ is, the more cancellations are required.
Does it really happen? This is the possible inconsistency mentioned above.
Similar questions can be asked about baryon-baryon scattering as well.

The examination of this problem led to some very interesting discoveries
recently \cite{JL,CC,MP}. By looking at the cases of $n=2$ and $n=3$,  it was
found
that many new and old quark-model-like relations must be obeyed, and
that an infinite tower of baryon resonances with certain characteristics
must be present. There are several ways to see the origin of these relations,
one of them is to regard the cancellation requirement as posing constraints
that are satisfied by these relations \cite{JL}. From that point of view, the
larger
$n$ is, the more must be cancelled and more constraints have to be satisfied.
So there is even a hope of getting more of these interesting relations by
understanding the cancellation mechanism for large $n$.

On the other hand, very little calculable dynamics is
put into large-$N_c$ QCD, so these cancellations are quite
likely of kinematical
origin peculiar to large-$N_c$ QCD, and the predictions may be mostly
a result of the Hartree nature of large-$N_c$ baryons \cite{GH,CC}.
We shall give further support to this point of view in the present and the
subsequent paper \cite{CS}.

The most distinct kinematical feature of a large-$N_c$ baryon is its
heavy mass, being of the order of $N_c$.
If a kinematical formula can be found to effect the cancellation in this case,
it seems likely that it can also be used in similar kinematical situations,
such as the presence of a heavy quark, or fermion-fermion scattering
at high energy in which the fermions carry large energy scale. We shall
return to these other interesting problems in the future.

We report in this paper a formula for the sum of the $n!$
diagrams, which can be used for these cancellations. We shall refer to it
as the {\it multiple commutator formula} for reasons to be mentioned later.
The formula and its predecessors can be formulated as exact combinatorial
theorems and for the sake of clarity we will concentrate on these mathematical
results in the present paper. Physical applications require
approximations at the appropriate kinematical regimes
and they will be deferred to future publications \cite{CS,YJ}.

This formula can be considered as a generalization of the `eikonal formula'
\cite{HC} useful in the discussion of soft-photon emission \cite{DR} and the
eikonalization
of high-energy scattering processes \cite{HCT}. The eikonal formula
(corollaries
1.1 and 2.1 in section 2) indeed gives a result for the sum of $n!$ diagrams,
but it is suitable only for vertices that are effectively
numbers. For pion-baryon problem in large-$N_c$ QCD, and high-energy
quark-quark scattering in usual QCD, strong-isospin and color are respectively
involved, so the vertices
are non-abelian and do not commute with one another. To deal with
the sum of diagrams in these
problems we need a non-abelian generalization of the eikonal formula,
which is the multiple commutator formula mentioned above.
It is considerably more complicated than the eikonal formula, but in the
limit of abelian vertices it reduces to the eikonal formula as it should.

The multiple commutator formula can be derived from a generalization
of the eikonal formula, to be called the factorization formula (theorems
1 and 2 of section 2). From that, we derive an intermediate `folding relation'
(theorems 3 and 4), which will be used to derive
the multiple commutator formula (theorems
5 and 6). Each of these relations have an on-shell version (theorems 2,3,5)
and an off-shell version (theorems 1,4,6). These two versions have identical
structures, and in fact the on-shell amplitudes can be obtained in a rather
surprising way from the off-shell amplitudes via a `domino relation' (theorem
7). The statements of these theorems, as well as illustrative examples,
will be presented in section 2, and their general proofs will be given in
section 3.

The on-shell version of the eikonal formula is in some sense a cancellation
formula. It expresses the sum of abelian amplitudes as a product of
$n$ energy-$\delta$-functions. In other words, sums of products of propagators
are replaced by $\delta$-functions.
In high-energy electron-electron scattering via $n$-photon exchange,
individual diagrams go like $s\ln(s)$ for large center-of-mass energy
$\sqrt{s}$, but this leading term is cancelled  in
the sum of the $n!$ diagrams to yield a final result going only like $s$
\cite{HCT}.
{}From the point of view of the eikonal formula, this cancellation
has to do with the
replacement of the propagators by the $\delta$-functions. When it
comes to high-energy quark-quark scattering in QCD, the eikonal formula
is no longer applicable, and indeed the $s\ln(s)$ contributions are
not cancelled, though explicit calculations show that they are concentrated
in the channel of color-octet exchange \cite{HCT}. In this non-abelian
situation, the
multiple commutator formula replaces the  eikonal formula, the sum of
the $n!$ non-abelian amplitudes is now given as sums of products of
energy-$\delta$-functions and multiple commutators of the vertices.
The term that contains pure $\delta$-functions is like the eikonal formula
and will lead to the same high-energy cancellations. On the other extreme,
there is a term that contains purely multiple commutators without
$\delta$-fucntions. This term leads to no reduction in the high energy
behavior but the quantum number of the exchange is a pure color octet
\cite{YJ}.

In contrast to the case of high energy scattering, the cancellation
needed in the large-$N_c$ baryon problem comes from the pure multiple
commutator term in the formula. It turns out that each commutator reduces
the power of $N_c$ by one \cite{CC,CS}, so the term with $n-1$ multiple
commutators
indeed reduces the individual amplitude by a factor $N_c^{n-1}$, as
required for consistency.

To the extent that multiple commutators are involved in the formula,
this non-abelian generalization of the eikonal formula bears some superficial
resemblance to formulas used in soft-pion productions when current algebra
is used \cite{SW}.
However, the similarity stops there, in that our formula is exact,
and is combinatorial and kinematical in nature. No dynamical assumptions
are involved, certainly not chiral symmetry nor the presence of a current
algebra. The kinematical regime that makes the formula useful
is the smallness of the boson energy compared to the mass of the fermion,
or the energy it carries, and not compared to some dynamical condensates like
$f_\pi$ as is the case in $\sigma$ models.
Of course our formula can be used in the soft-pion situation
as well, but formulas developed there are not general enough for our
purposes.

With this brief mention of physical examples to illustrate what the formulas
can be
used for, we will devote the rest of the paper to purely mathematical
results, to explain, illustrate, and prove the stated theorems.

\section{Combinatorial Formulas}
\subsection{ Trees and their amplitudes:}

\begin{figure}
\vskip -4 cm
\centerline{\epsfxsize 4.7 truein \epsfbox {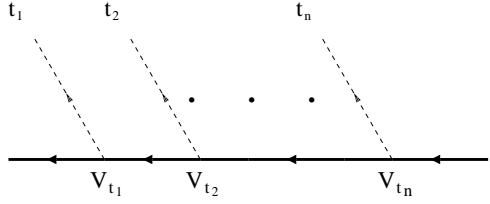}}
\nobreak
\vskip -9.5 cm\nobreak
\vskip .1 cm
\caption{ A standard tree diagram $[t_1t_2\cdots  t_n]$.}
\end{figure}

 It is necessary to establish relatively simple notations to discuss
these rather complicated combinatorial problems. To that end,
let us denote a tree diagram  arranged in the manner of
 Fig.~1 as
$T=[t_1t_2\cdots   t_n]$.
A tree $T=[o]$ with a single boson is just a vertex.
If tree $T_1=[t_{11}\cdots   t_{12}]$ is joint to the left of tree
$T_2=[t_{21}\cdots   t_{22}]$, the resulting tree $[t_{11}\cdots
t_{12}t_{21}\cdots  t_{22}]$
will be denoted by $[T_1T_2]$ or $[T_1|T_2]$.
Note that $[T\emptyset]=[T|\emptyset]=T$ for a null tree
 $\emptyset$.

Given a number of trees $T_i$,  ${\cal T}=\{T_1;T_2;T_3;\cdots  \}$ is defined
to be
the set of all trees obtained by merging them together in all possible
ways. If $N_i=N[T_i]$ is the length (number of boson lines)
 in tree $T_i$, then the number of trees in the set ${\cal T}$ is
given by the multinomial coefficient
\begin{eqnarray}
C(N_1,N_2,N_3,\cdots )=
(\sum_iN_i)!/\prod_iN_i!\ ,
\end{eqnarray}
and the length of each tree is $\sum_iN_i$.
For example, if $T_1=[12]$ and $T_2=[345]$, then
${\cal T}=\{T_1;T_2\}=\{12;345\}$ is the set of 10 trees
$[12345]$, [13245], [13425], [13452],
[31245], [31425], [34125], [31452], [34152], [34512].
In particular, $\{t_1;t_2;\cdots ;t_n\}$ is the set of $n!$ trees
obtained by permuting the boson lines in the tree $T=[t_1t_2\cdots t_n]$.
We shall adopt the notation $\{T;\}$ to denote this permutation set of $T$.

If we join an extra tree $T_e$ to the right of every tree in set ${\cal T}$,
the
resulting set of trees is denoted by $\{T_1;T_2;T_3;\cdots |T_e\}$, or
simply $\{{\cal T}|T_e\}$.
The number of trees in this set is the same as in ${\cal T}$,
but the length of each tree is $(\sum_iN_i)+N_e$.
In the example above,
if $T_e=[67]$,
then $\{T_1;T_2|T_e\}=\{12;345|67\}$ consists of the ten trees
$[1234567]$,
[1324567], $\cdots $,  [3415267], [3451267].

Given a tree $T=[t_1\cdots  t_n]$, the `energy' of boson line $t_i$ in the rest
frame of the fermion will be denoted by $\omega_{t_i}$. The `total energy'
of the tree is
\begin{eqnarray}
W[T]=\sum_{i=1}^n \omega _{t_i}\
\end{eqnarray}

The {\it off-shell `amplitude'} of a tree
$T=[t_1t_2\cdots  t_n]$ is {\it defined} to be the product of $n$
`propagators',
\begin{eqnarray}
a^*[T]=a^*[t_1t_2\cdots  t_n]=\prod_{p=1}^n{1\over\sum_{j=1}^p\omega_{t_j}
+i\epsilon}\ ,
\end{eqnarray}
and the {\it on-shell `amplitude'} of the same tree is obtained
by replacing the last `propagator' by an `energy-conserving'
$\delta$-function:
\begin{eqnarray}
a[T]=a[t_1t_2\cdots  t_n]=-2\pi
i\delta\left(\sum_{j=1}^n\omega_{t_j}\right)
\prod_{p=1}^{n-1}{1\over\sum_{j=1}^p\omega_{t_j} +i\epsilon}\
{}.
\end{eqnarray}
If the tree $T=[o]$ has only one line, then
by definition we let $a[o]=-2\pi i\delta(\omega_o)$. Also,
$a[To]=-2\pi i\delta\left(W[To]\right)a^*[T]$ for any $T$.

Incidentally, we put quotes on words like `amplitude', `propagator', and
`energy' to avoid confustion when these formulas are used in cases
when they are not {\it exactly} the amplitude, propagator, and energy of the
physical process on hand. The theorems
proved here are combinatorial and mathematical and the actual physical meaning
of quantities like $a^*, a$ and $\omega_{t_i}$ are irrelevant.
However, it is useful to use these names to give them some physical intuition.

It will also be convenient to adopt the notation $a^*_E[T]$ to mean the
off-shell
amplitude with `energy' amount $E$ added to
the first boson line $t_1$ of $T$:
\begin{eqnarray}
a^*_E[T]=a^*_E[t_1t_2\cdots  t_n]=\prod_{i=1}^n{1\over
E+\sum_{j=1}^i\omega_{t_j}
+i\epsilon}\ .
\end{eqnarray}
The modified on-shell amplitude $a_E[T]$ can be defined similarly.
In this notation, it is clear that
\begin{eqnarray}
a^*[T_1T_2]=a^*[T_1]a^*_{W[T_1]}[T_2]\ ,\quad
a[T_1T_2]=a^*[T_1]a_{W[T_1]}[T_2] \ .
\end{eqnarray}

The off-shell amplitude for a set is defined to be
the sum of the amplitudes for the individual trees in the set. Thus
\begin{eqnarray}
a^*\{{\cal T}\}=\sum_{T\in {\cal T}}a^*[T]\ ,
\end{eqnarray}
and
\begin{eqnarray}
a^*\{{\cal T}|T_e\}=\sum_{T\in {\cal T}}a^*[TT_e]\ .
\end{eqnarray}
On-shell amplitudes $a\{{\cal T}\}$ and $a\{{\cal T}|T_e\}$ for these sets are
defined
similarly.

\subsection{ Main Theorems:}

We can now state the main combinatorial results, whose proofs are to be found
in Section 3.
The first result is a
{\it factorization formula} for off-shell amplitudes:

\noindent{\it Theorem 1 (off-shell factorization formula):}\quad
\begin{eqnarray}
a^*\{T_1;T_2;T_3;\cdots \}=\prod_i a^*[T_i]\ .
\end{eqnarray}

In the special case when every tree $T_i=[t_i]$ is
just a single vertex, $\{T;\}=\{t_1;t_2;\cdots ;t_n\}$ is the set of $n!$
permuted trees of $T=[t_1t_2\cdots t_n]$. Then (2.9) reduces to a known
formula which we shall refer to as  the `off-shell abelian eikonal formula':

\noindent{\it Corollary 1.1 (off-shell abelian eikonal formula):}
\begin{eqnarray}
a^*\{t_1;t_2;\cdots ;t_n\}=\prod_{i=1}^na^*[t_i]\ .
\end{eqnarray}

As is well known, this formula can be used for example to show that soft
photons emitted from a charged particle satisfies the Poisson distribution.

\noindent{\it Example 1:}\quad
Suppose
$T_1=[1]$ and $T_2=[23]$. Then (2.9) reads
$a^*\{1;23\}=a^*[1]a^*[23]$. Explicitly, this says
\begin{eqnarray}
&&{ 1\over{\omega _1(\omega _1+\omega _2)
(\omega _1+\omega _2+\omega _3)}}+
{1\over{\omega _2(\omega _2+\omega _1)
(\omega _2+\omega _1+\omega _3)}}+
{ 1\over {\omega _2(\omega _2+\omega _3)
(\omega _2+\omega _3+\omega _1)}}\nonumber\\
&=&
{ 1\over{\omega _1 \omega _2(\omega _2+\omega _3)}} \ ,
\end{eqnarray}
an algebraic identitiy that can be verified directly.
For simplcity in printing, we have omitted the $+i\epsilon$ which should
appear in all the denominators.

There is also a factorization formula for on-shell amplitudes. It states

\noindent{\it Theorem 2 (on-shell factorization formula):}
\begin{eqnarray}
a\{T_1;T_2;\cdots \}=\prod_ja[T_j]\
 .
\end{eqnarray}

Note that the right-hand side vanishes if the energies are generic,
{\it viz.,} when $W[T_i]\not=0$ for any $i$, though
of course by definition we require $\sum_iW[T_i]=0$. Thinking of it that
way, this theorem tells us what on-shell amplitudes can be added up to
cancel one another. Even when the energies are not generic, the presence
of additional energy-conserving $\delta$-functions on the right also lead to
partial
cancellations, at least in high-energy scattering amplitudes. See the remark
below Corollary 2.1.

When every tree $T_i=[t_i]$ is just a single vertex, this formula reduces
to the `on-shell abelian eikonal formula':
\noindent{\it Corollary 2.1 (on-shell abelian eikonal formula):}
\begin{eqnarray}
a\{t_1;t_2;\cdots ;t_n\}=\prod_{j=1}^na[t_j]=
\prod_{j=1}^n\left(-2\pi i\delta(\omega_j)\right)\ .
\end{eqnarray}

This formula has been used to obtain large cancellations between
Feynman diagrams in high-energy electron-electron
scattering, resulting thereby an eikonal scattering formula in the
sum. The main goal of this paper is to obtain a non-abelian generalization
for this formula.

We illustrate this theorem with a
simple example.

\noindent{\it Example 2:}\quad Consider the same trees as in Ex.~1.
For these trees,
eq.~(2.12) demands $a\{1;23\}=\left(-2\pi
i\right)^2\delta(\omega_1+\omega_2+\omega_3)
\delta(\omega_1)(\omega_2+i\epsilon)^{-1
}$. This identity can be directly verified by using
$W_{tot}=\omega_1+\omega_2+\omega_3=0$:
\begin{eqnarray}
&&{1\over{(\omega _1+i\epsilon)(\omega _1+\omega _2+i\epsilon)}}+
{ 1\over{(\omega _2+i\epsilon)(\omega _2+\omega _1+i\epsilon)}}+
{ 1\over{(\omega _2+i\epsilon)(\omega _2+\omega _3+i\epsilon)}}\nonumber\\
&=&{ 1\over{\omega_2+i\epsilon}}\left({ 1\over{\omega_1+i\epsilon}}+
{1\over{-\omega_1+i\epsilon}}\right)=-2\pi i\delta(\omega_1){
1\over{\omega_2+i\epsilon}}\
{}.
\end{eqnarray}

We need more notations to go on. The
transpose of a tree $T=[t_1t_2\cdots t_{n-1}t_n]$ is defined to be
the tree $\tilde T=[t_nt_{n-1}\cdots t_2t_1]$.  A subtree of $S=[s_1\cdots
s_N]$
from $s_p$ to $s_q$ will be denoted by $\sigma_{p,q}=[s_ps_{p+1}\cdots s_q]$;
thus
$S=\sigma_{1,N}=[\sigma_{1,k}\sigma_{k+1,N}]$ for
any $k$.

We will consider the tree $[RoS]$
obtained by joining line $o$ to the left of
tree $S$, and then $R$ to the left of the tree $[oS]$. This tree
is of length $N[R]+1+N[S]\equiv N'+1+N$.
The next combinatorial formula enables the
on-shell amplitude of this tree to be expressed as combinations of on-shell
amplitudes in which line $o$ is at the extreme right. This result is needed
to obtain the non-abelian generalization of the eikonal formula
(2.13).

\noindent{\it Theorem 3 (on-shell folding formula):}
\begin{eqnarray}
a[RoS]=\sum_{k=0}^N(-)^k
a\{R;
\tilde \sigma_{1,k}|o\}a[\sigma_{k+1,N}]\ .
\end{eqnarray}
For $k=N$, the tree $\sigma_{N+1,N}=\emptyset$ is null, and we should
interpret the factor $-2\pi i\delta\left(W[\emptyset]\right)$ simply as 1.

We call this formula the {\it folding formula} because
it can be obtained by a folding operation.
Consider first the term $k=N$. This can be obtained by folding the linear tree
$[RoS]$ (Fig.~2a) at the point
$o$ (Fig.~2b) and merging together the resulting trees
$R$ and $\tilde S$ in all possible ways
to form $a\{R;\tilde S|o\}$ (Fig.~2c). For terms with $k<N$, we must
tear off the subtree $\sigma_{k+1,N}$ at the end before folding and merging.

Since $a\{R;\tilde\sigma_{1,k}|o\}=-2\pi
i\delta\left(W[R\tilde\sigma_{1,k}o]\right)
a^*\{R;\tilde\sigma_{1,k}\}$, with the
help of (2.9), we may write (2.15) as

\noindent{\it Corollary 3.1:}
\begin{eqnarray}
a[RoS]=
-2\pi i\sum_{k=0}^N(-)^k\delta\left(W[R\tilde{\sigma_{1,k}}o]\right)
a[\sigma_{k+1,N}]a^*[R]a^*[\tilde \sigma_{1,k}]\ .
\end{eqnarray}

\begin{figure}
\vskip -8 cm
\centerline{\epsfxsize 5 truein \epsfbox {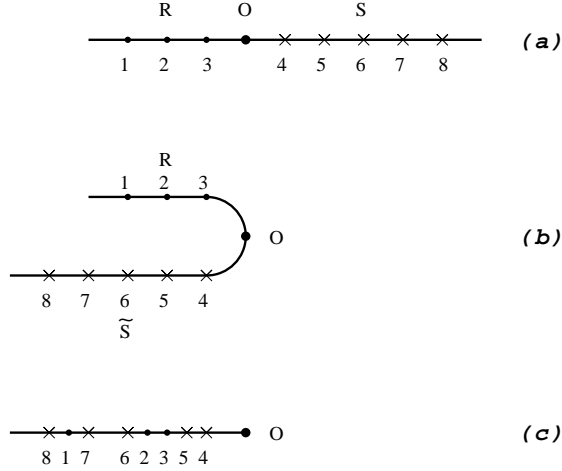}}
\nobreak
\vskip 5 cm\nobreak
\vskip .1 cm
\caption{Folding operation for theorems 3 and 4. }
\end{figure}

\noindent{\it Example 3:}\quad
Suppose $R=[1]$ and $S=[23]$.
Then (2.16) demands
\begin{eqnarray}
a[1o23]=\{-2\pi i\delta(\omega_2+\omega_3)a[23]a^*[1]
+2\pi i\delta(\omega_3)a^*[1]a^*[2]+a^*[1]a^*[32]\}\Delta\ ,
\end{eqnarray}
where $\Delta\equiv -2\pi i\delta(\omega_1+\omega_o+\omega_2+\omega_3)$.
To verify this directly, we use `energy conservation' $\omega_1+\omega_o+
\omega_2+\omega_3=0$ to rewrite the left-hand side of (2.17) as
\begin{eqnarray}
a[1o23]&=&{1\over{(\omega_1+i\epsilon)(\omega_1+\omega_o+i\epsilon)
(\omega_1+\omega_o+\omega_2+i\epsilon)}}\Delta\nonumber\\
&=&{1\over{(\omega_1+i\epsilon)(-\omega_2-\omega_3+i\epsilon)
(-\omega_3+i\epsilon)}}\Delta\nonumber\\
&=&{1\over{\omega_1+i\epsilon}}\left({
1\over{(\omega_2+\omega_3+i\epsilon)}}+2\pi i\delta
(\omega_2+\omega_3)\right)\left(
{1\over{\omega_3+i\epsilon}}+2\pi i\delta(\omega_3)\right)\Delta\ ,
\end{eqnarray}
which gives (2.17).

There is an off-shell extension of the folding theorem:

\noindent{\it Theorem 4 (off-shell folding formula):}
\begin{eqnarray}
a^*[RoS]&=\sum_{k=0}^{N}(-)^{k}a^*
\{R;\tilde
\sigma_{1,k}|o\}a^*[\sigma_{k+1,N}]\ .&
\end{eqnarray}
($a^*[\emptyset]\equiv 1$).
{}From (2.6) and (2.9),
$a^*\{R;\tilde \sigma_{1,k}|o\}=
a^*[R]a^*[\tilde\sigma_{1,k}]\left(W[R\tilde\sigma_{1,k}o]+i \epsilon \right)
^{-1}$. Hence (2.19) can also be written as

\noindent{\it Corollary 4.1:}
\begin{eqnarray}
a^*[RoS]=\sum_{k=0}^{N}(-)^k
a^*[R]a^*[\tilde\sigma_{1,k}]a^*[\sigma_{k+1,N}]/\left(W[R\tilde
\sigma_{1,k}o]+i\epsilon\right)\
{}.
\end{eqnarray}

\noindent{\it Example 4:}\quad
Using again $R=[1]$ and $S=[23]$, (2.20) demands
\begin{eqnarray}
a^*[1o23]=a^*[1]
\left({{a^*[23]}\over{\omega_1+\omega _o}}-{{a^*[2]a^*[3]}\over{
\omega _1+\omega _2+\omega _o}}+{{a^*[32]}\over{
\omega _1+\omega _3+\omega _2+\omega _o}}\right)\ .
\end{eqnarray}
Explicitly, this requires
\begin{eqnarray}
&&{ 1\over{\omega _1(\omega _1+\omega _o)(\omega _1+\omega _o+
\omega _2)(\omega _1+\omega _o+
\omega _2+\omega _3)}}\nonumber\\
&=&{1\over{\omega_1}}\left({ 1\over{\omega _2(\omega _2+\omega _3)
(\omega _1+\omega _o)}}-{ 1\over{\omega _2\omega _3
(\omega _1+\omega _2+\omega _o)}}\right)\nonumber\\
&+&{1\over{\omega_1}}\left({ 1\over{\omega _3(\omega _3+\omega _2)
(\omega _1+\omega _3+\omega _2+\omega _o)}}\right)\ ,
\end{eqnarray}
which can be directly verified to be true. Again, we have omitted
the $+i\epsilon$ which should appear everywhere in the denominators.

We proceed to discuss non-abelian amplitudes. In this case the vertices
$V_i$ are matrices that do not commute with one another. We shall define
the operator
$V[T]$ for  a tree $T=[t_1t_2\cdots t_n]$ to be the product of its vertices,
\begin{eqnarray}
V[T]=V_{t_1}V_{t_2}\cdots V_{t_n}\ ,
\end{eqnarray}
and the multiple commutator of the tree to be
\begin{eqnarray}
V^{MC}[T]=[V_{t_1}[V_{t_2}[\cdots [V_{t_{n-1}},V_{t_n}]\cdots ]]]\ .
\end{eqnarray}
If the tree $T=[t]$ is a single vertex, $V^{MC}[t]$ is taken to mean $V[t]
=V_t$.

Recall that $\{T;\}=\{t_1;t_2;\cdots ;t_n\}$ is the set of $n!$ trees obtained
by permuting the lines of a single tree $T=[t_1t_2\cdots t_n]$. The {\it
`non-abelian on-shell amplitude'} is defined to be
\begin{eqnarray}
\sum_{K\in\{T;\}}a[K]V[K]\equiv aV\{T;\}\ .
\end{eqnarray}
The off-shell non-abelian amplitude $a^*V\{T;\}$ is defined similarly.
Our aim is to find a generalization of (2.13) suitable for the non-abelian
amplitudes.
Naturally for on-shell amplitudes this generalization must reduce to (2.13)
when all $V_{t_i}=1$.

To state the result we must first  establish some convenient ordering of the
numbers $t_i$ appearing in the trees of $\{T;\}$. We shall choose the
natural order of the original tree $T$, namely,
$t_1<t_2<\cdots <t_n$. In terms of this ordering, a partition will be
established for every tree $K$ of $\{T;\}$
in the following way. If the rightmost element of $K$ is larger than
any other element to its left, then there is one partition and it is the
whole tree $K$. Otherwise, come to the first number larger than it and
draw a partition just before that number. Now start again just to the left
of the new partition and repeat the same procedure. If this number is
now larger than all others to its left, there will be no more partitions.
Otherwise another partition will appear just before the first number
encountered which is larger than it. Repeat this procedure until the
whole tree $K$ is partitioned.

For example, suppose $T=[1234567]$ so the ordering is $1<2<3<4<5<6<7$.
The partitions of the following sample trees $K$ are shown to the right of
the arrow: $[1234567] \to [1234567]$, $[1437562] \to [14|37|56|2]$,
$[1357246] \to [1357|246]$, $[7654321] \to [7|6|5|4|3|2|1]$.

Let $K=[K_1|K_2|\cdots |K_m]$ be the partition of a tree in $\{T;\}$.
Then

\noindent{\it Theorem 5 (on-shell multiple commutator formula):}
\begin{eqnarray}
aV\{T;\}&\equiv &\sum_{K\in\{T;\}}a[K]V[K]\nonumber\\
&=&\sum_{K\in\{T;\}}\left(\prod_{a=1}^m a[K_a]\right) V^{MC}[K_1]
V^{MC}[K_2]\cdots V^{MC}[K_m]\ .
\end{eqnarray}

In the abelian limit when $V_{t_i}=1$ for every $i$, $V^{MC}[K_a]=0$
unless $K_a$ is a single vertex $[t_a]$, whence $V[K_a]=V[t_a]=1$.
The only tree in $\{T;\}$ that fulfils this condition
is the tree $\tilde T
=[t_nt_{n-1}\cdots t_2t_1]$, whose partition is $[t_n|t_{n-1}|\cdots
|t_2|t_1]$.
This is the only tree that contributes to the right of (2.26)
and its contribution is $\prod_{i=1}^na[t_i]$. In the present case the
left-hand side of (2.26)
is $a\{T;\}=a\{t_1;t_2;\cdots ;t_n\}$, so eq.~(2.26) reduces
to (2.13).

For generic energy configurations, where the sum of all energies $\omega_{t_i}$
vanishes but the sum of any subset does not, $a[K_i]=0$ unless $K_i$ is the
whole tree $K$. These trees with one partition are precisely those
with $t_n$ at the end, and there are $(n-1)!$ of them.
Using the notation $T'$ to denote the tree
$[t_1t_2\cdots t_{n-1}]$, and $\{T';\}$ to denote its permutation set of
$(n-1)!$
trees, one gets

\noindent{\it Corollary 5.1:}\quad {For generic energy configurations,}
\begin{eqnarray}
aV\{T;\}=\sum_{J\in\{T';\}}a[Jt_n]V^{MC}[Jt_n]\ .
\end{eqnarray}

This corollary can be used to establish the consistency of
scattering in the large-$N_c$ limit. Take for example the case of $n$ mesons
scattered from a baryon. It is known that for $N_c\gg 1$,
individual tree diagrams grow   like $N_c^{n/2}$ but the sum of the $n!$
diagrams must decline like $N_c^{1-n/2}$. This calls for
a cancellation of $n-1$ powers of $N_c$ in the sum! It is shown elsewhere
\cite{CS}
that eq.~(2.27) can be used to establish this cancellation.

For QCD as well as QED scattering amplitudes in the high-energy limit,
dominant contributions often get cancelled between diagrams. Theorem 5
can also be used to compute such cancellations.

\noindent{\it Example 5.1:}\quad We will use (2.15) to compute directly
$aV\{T;\}$ for $T=[12]$. According to that formula,
$a[21]=a[1]a[2]-a[12]$. Hence
\begin{eqnarray}
aV\{T;\}=a[12]V_1V_2+a[21]V_2V_1=a[12][V_1,V_2]+a[1]a[2]V_2V_1\ ,
\end{eqnarray}
which agrees with (2.26).

\noindent{\it Example 5.2:}\quad We proceed to compute the more difficult case
of $T=[123]$. Then
\begin{eqnarray}
aV\{T;\}&=a[123]V_1V_2V_3+a[132]V_1V_3V_2+a[213]V_2V_1V_3&\nonumber\\
&+a[231]V_2V_3V_1+a[312]V_3V_1V_2+a[321]V_3V_2V_1\ .&
\end{eqnarray}
Using (2.15), one gets
\begin{eqnarray}
a[132]&=&a[13]a[2]-a[1;2|3]=a[13]a[2]-a[123]-a[213]\ ,\nonumber\\
a[231]&=&a[23]a[1]-a[2;1|3]=a[23]a[1]-a[213]-a[123]\ ,\nonumber\\
a[312]&=&a[3]a[12]-a[1|3]a[2]+a[21|3]=a[3]a[12]-a[13]a[2]+a[213]\ ,\nonumber\\
a[321]&=&a[3]a[21]-a[2|3]a[1]+a[12|3]=a[3]a[21]-a[23]a[1]+a[123]\nonumber\\
&=&a[3]\left(a[1]a[2]-a[12]\right)-a[23]a[1]+a[123]\ .
\end{eqnarray}
Substituting (2.30) into (2.29) and grouping the terms into multiple
commutators, one gets
\begin{eqnarray}
&a[123][V_1[V_2,V_3]]+a[213][V_2[V_1,V_3]]+a[3]a[12]V_3[V_1,V_2]&\nonumber\\
+&a[13]a[2][V_1,V_3]V_2+a[23]a[1][V_2,V_3]V_1+a[3]a[2]a[1]V_3V_2V_1\
{}.
\end{eqnarray}
This agrees with (2.26) because the correct partitions of the six
trees in $\{T;\}$ are $[123] \to [123]$, $[213] \to [213]$,
$[312] \to [3|12]$, $[132] \to [13|2]$,
$[231] \to [23|1]$,  $[321]
\to [3|2|1]$.

There is an off-shell version of Theorem 5:

\noindent{\it Theorem 6 (off-shell multiple commutator formula):}
\begin{eqnarray}
a^*V\{T;\}&\equiv& \sum_{K\in\{T;\}}a^*[K]V[K]\nonumber\\
&=&\sum_{K\in\{T;\}}\left(\prod_{a=1}^m a^*[K_a]\right) V^{MC}[K_1]
V^{MC}[K_2]\cdots V^{MC}[K_m]\ .
\end{eqnarray}

We have discussed three off-shell theorems and three on-shell theorems,
and in each case the off-shell relations are identical to the on-shell
relations.
This is somewhat surprising because left-hand side of these relations is linear
in the amplitude, but the right-hand side is not. Starting from an
off-shell amplitude, we can replace its
last propagator by an energy-conservation $\delta$-function to turn it into an
on-shell amplitude. When this is done on the left-hand side,
somehow  the off-shell amplitudes on the right must all turn into on-shell
amplitudes as well, in spite of the non-linearity appearing
on the right. This domino effect can be expressed as

\noindent{\it Theorem 7 (domino relation):}
\begin{eqnarray}
 2\pi\delta\left(\sum_iW[T_i]\right)\lim_{\epsilon\to
0}\left(N\epsilon\prod_{i}a^*[T_i]\right)=\prod_{i}a[T_i]\ ,
\end{eqnarray}
where $N=\sum_iN[T_i]$ is the total number of boson lines in all the trees.

To demonstrate this relation we must pay special care to the $i\epsilon$'s.
As they are defined in (2.3) and (2.4) it is meant to be true when $\epsilon\to
0$. However, expressions like (2.9) and (2.11) are also true for
{\it finite} $\epsilon$'s provided an $i\epsilon$ is added to {\it each}
$\omega_i$. This means that we should really have written $ip\epsilon$
instead of $i\epsilon$ in (2.3) and (2.4). In particular,
the $i\epsilon$ associated with the last
propagator of $a\{T_1;T_2;\cdots ;\}$ is actually $iN\epsilon$.
With this in mind, the left-hand side of (2.9) is turned into an on-shell
amplitude by multiplying it by
\begin{eqnarray}
-2\pi i\delta\left(\sum_iW[T_i]\right)
(\sum_iW[T_i]+iN\epsilon)=2\pi N\epsilon\delta\left(\sum_iW[T_i]\right)\
{}.
\end{eqnarray}
When this factor is multiplied to both sides of (2.9) and the result compared
with (2.12), Theorem 7 is obtained.

\noindent{\it Example 7:}\quad We demonstrate explicitly here the special
case $T_i=[t_i]$, so $\{T_1;T_2;\cdots \}=\{t_1;t_2;\cdots ;t_n\}$.
To obtain (2.33), we need to verify that
\begin{eqnarray}
\lim_{\epsilon\to 0}(in\epsilon)\left(\prod_{j=1}^{n-1}{
1\over{\omega_j+i\epsilon}}\right)
\left({
1\over{-\sum_{j=1}^{n-1}\omega_j+i\epsilon}}\right)=
\prod_{j=1}^{n-1}\left[-2\pi
i\delta(\omega_j)\right] \ .
\end{eqnarray}
One way of proving it is to take any function
$f(\omega_1,\omega_2,\cdots ,\omega_{n-1})$, analytic in the lower $\omega_j$
planes, and vanishs sufficiently fast at infinity to allow the integration of
\begin{eqnarray}
(-in\epsilon)\int_{-\infty}^\infty \prod_{j=1}^{n-1}\left(\o
1 {\omega_j+i\epsilon}d\omega_j\right)
f(\omega_1,\omega_2,\cdots ,\omega_{n-1}){
1\over{\sum_{j=1}^{n-1}\omega_j-i\epsilon}}
\end{eqnarray}
to be carried out by completing contours in the lower-half planes.
The presence of the factor $F=(\sum_{j=1}^{n-1}\omega_j-i\epsilon)^{-1}$
produces a pinch which ensures the integral not to vanish. The result of these
$(n-1)$ integrations is to produce a factor $(-2\pi i)^{n-1}$
and to put every $\omega_j=-i\epsilon$, whence $F=(-in\epsilon)^{-1}$,
so (2.33) is immediately seen to be true.

\setcounter{equation}{0}
\section{ Proofs}

The theorems stated in Section 2 will be proved here.

\subsection{ Proof of Theorem 1}

The easiest way to prove it is to use the Schwinger representation to write the
off-shell amplitude for a tree $T=[t_1t_2\cdots t_n]$ as
\begin{eqnarray}
a^*[T]&=(-i)^n\int_Td^n\tau\exp\left(i\sum_{i=1}^n\omega_{t_i}
\tau_{t_i}\right)&\nonumber\\
\int_Td^n\tau&\equiv\int_0^\infty d\tau_{t_n}\int_{\tau_{t_n}}^\infty
d\tau_{t_{n-1}} \cdots \int_{\tau_{t_{2}}}^\infty d\tau_{t_1}\ .&
\end{eqnarray}
An integration variable $\tau_{t_i}$ has been introduced for every boson line,
and these variables are ordered
($\infty>\tau_{t_1}\ge\tau_{t_2}\ge\cdots \ge\tau_{t_n}\ge 0$) precisely the
same
way  the boson lines in the tree $T=[t_1t_2\cdots  t_n]$ are ordered.

Given this relation between the integration region and the ordering of
the (boson lines in the) tree, eq.~(2.9) is almost obvious. For any $T\in{\cal
T}$,
the ordering of integration variables in $a^*[T]$ is determined by
the ordering of $T\in{\cal T}$.
Thus the variables within
each tree $T_i$ are ordered by those individual trees, whatever $T\in{\cal T}$
is.
The relative orderings
of the variables for different $T_i$'s, however, depend on $T$, but when
summed over all $T\in{\cal T}$, every possible relative orderings are included,
so
the union of the integration regions factorizes into
the product of the integration regions of $[T_i]$
for different $i$'s. Hence (2.9) is proved.

\subsection{ Proof of Theorem 2}

The proof is almost identical to that of Theorem 1. Instead of (3.1),
the on-shell amplitude $a[T]$ is given by
\begin{eqnarray}
-2\pi i\delta(W[T])a[T]
&=(-i)^n\int_{T_o}d^n\tau\exp\left(i\sum_{i=1}^n\omega_{t_i}
\tau_{t_i}\right)&\nonumber\\
\int_{T_o}d^n\tau&\equiv\int_{-\infty}^\infty
d\tau_{t_n}\int_{\tau_{t_n}}^\infty
d\tau_{t_{n-1}} \cdots \int_{\tau_{t_{2}}}^\infty d\tau_{t_1}\ .&
\end{eqnarray}
This differs from (3.1) only in the integration region $T_o$, which is
now defined by the ordering $\infty> \tau_{t_1}\ge \tau_{t_2} \ge \cdots  \ge
\tau_{t_n}> -\infty$. The rest of the argument goes similarly. When summed
over all $T\in{\cal T}=\{T_1;T_2; \cdots \}$, the integration variables
$\tau_{t_a}$
retains only the ordering within each individual tree $T_i$, and for each
tree they integrate from $-\infty$ to $+\infty$. Using (3.2) again for
indiviual trees $T_i$, we obtain (2.12).

\subsection{ Proof of Theorem 3}

The proof can be established by induction on the length $N=N[S]$ of the
tree $S$. If $N=0$, (2.15) is trivially true. Now assume the formula
to be true up to $N=m-1$ for an arbitrary tree $R$,
We must show it to be true also for $N=m$.

Let $S=[s_1s_2\cdots s_m]$. It will be useful in the following to divide
this into two subtrees $S=[\sigma_{1,k}\sigma_{k+1,m}]$, with
$\sigma_{p,q}=[s_{p}\cdots s_q]$. Let $T_1=[Ro], T_2=[S]$, and
${\cal T}=\{T_1;T_2\}$. From (2.12),
\begin{eqnarray}
a\{Ro;s_1s_2\cdots s_m\}=\sum_{T\in{\cal T}}a[T]=a[Ro]a[S]\ .
\end{eqnarray}
The sum over $T$ can be organized according to the number $k$ of boson lines
in $S$ appearing to the left of line $o$, as follows:
\begin{eqnarray}
\sum_{T\in{\cal T}}a[T]=a[RoS]+\sum_{k=1}^{m}a\{R;\sigma_{1,k}|o
\sigma_{k+1,m}\}\
{}.
\end{eqnarray}
For $k\ge 1$, there are at most $(m-1)$ lines to the right of $o$, so we can
use the induction hypothesis to write
\begin{eqnarray}
a\{R;\sigma_{1,k}|o\sigma_{k+1,m}\}=\sum_{\ell=0}^{m-k}(-)^{\ell}
a\{R;\sigma_{1,k};\tilde\sigma_{k+1,k+\ell}|o\} a[\sigma_{k+\ell+1,m}]  \
{}.
\end{eqnarray}
Substituting this into (3.4) and (3.3), and introducing the variable
$k'=k+\ell$, one recovers (2.15) for $N=m$ as follows:

\begin{eqnarray}
a[RoS]&=&a[Ro]a[S]-\sum_{k'=1}^m(-)^{k'}
a[\sigma_{k'+1,m}]\sum_{k=1}^{k'}(-)^k  a\{R;\sigma_{1,k};
\tilde\sigma_{k+1,k'}|o\}\nonumber\\
&=&a[Ro]a[S]+\sum_{k'=1}^m(-)^{k'}a\{R;\tilde
\sigma_{1,k'}|o\}a[\sigma_{k'+1,m}]\nonumber\\
&=&\sum_{k'=0}^m(-)^{k'}
a\{R;\tilde\sigma_{1,k'}|o\}a[\sigma_{k'+1,m}]\ .
\end{eqnarray}

In reaching this conclusion, we have used the relation
\begin{eqnarray}
\sum_{k=0}^{k'}(-)^ka\{R;\sigma_{1,k};\tilde\sigma_{k+1,k'}|o\}=0
\end{eqnarray}
which can be proven as follows. Define $a<b$ to mean line $a$ lying to the
left of line $b$ in a tree. Every tree in
${\cal T}_k\equiv\{R;\sigma_{1,k};\tilde\sigma_{k+1,k'}\}$ obeys the order
$s_1<s_2<\cdots <s_k\circ s_{k+1}>s_{k+2}>\cdots  >s_{k'}$, where $s_k\circ
s_{k+1}$
means that  both $s_k<s_{k+1}$ and $s_k>s_{k+1}$ are allowed.
In the former case, $(-)^ka\{{\cal T}_k\}$ is cancelled by a similar term in
$(-)^{k+1}a\{{\cal T}_{k+1}\}$, and in the latter case, it is cancelled by a
similar
term in $(-)^{k-1}a\{{\cal T}_{k-1}\}$, as long as $1\le k\le k'-1$. Moreover,
the
terms in $k=0$ and completely cancelled by terms in $k=1$, and terms in $k=k'$,
are completely cancelled by terms in $k=k'-1$. We have thus established
(3.7), hence (3.6) and the induction proof.

\subsection {Proof of Theorem 4}

The proof is identical to that of Theorem 3, with $a$ replaced by $a^*$,
and (2.15) replaced by (2.19).

\subsection{Proof of Theorem 5}

We shall refer to amplitude of the type $\prod_{i=1}^ma[K_i]$
appearing on the right of (2.26) as {\it derived amplitudes}. Each of its
factors $a[K_i]$ will be called an {\it allowed factor}. Given a subtree
$S$ of a tree in $\{T;\}$, its amplitude $a[S]$ is  an allowed factor
iff the last element of $S$ is larger than all its other elements.
A derived amplitude is a product of allowed factors for non-overlapping
subtrees.

There is a derived amplitude for each tree $K\in\{T;\}$, so all together there
are $n!$ distinct derived amplitudes. The amplitude $a[K]$ of
a tree is not equal to its derived amplitude, but by using the folding
formula (2.15) it can be expanded as a linear combination of the $n!$
derived amplitudes, with coefficients either $-1,0$, or $+1$.
To carry out the expansion, first locate the largest element $o$ in $K$ and use
(2.15) to fold
$a[K]$ about $o$.  This gives rise to sums of product of two amplitudes, the
first of which is always an allowed factor because of the choice of $o$ being
the largest element in $K$. The second amplitude in the product may not
be an allowed factor, but if it is not, we can locate its largest element
and use (2.15) to fold it about this element
again. Continuing thus, we can reduce $a[K]$ to a sum of products of
allowed factors, and hence a sum of derived amplitudes. Examples 5.1
and 5.2 in Section 2 contain explicit examples of how this is done.

We also need to know where every derived amplitude and its allowed factors
$a[K_a]$ come from. If they come from a tree $L$, then every $K_a$ is
obtained from a section of $L$ by folding over the largest
element in that section. To ask all allowed configurations of that section
is to ask all the possible ways to unfold $K_a$ about its end point.
This is essentially doing Fig.~2 in reverse, starting from the tree $K_a$
in Fig.~2c with its end point being $o$, then proceed to separate the rest
of the tree $K_a$ into two distinct parts $R$ and $\tilde S$ as in Fig.~2b,
and finally unfolding it into the tree $RoS\equiv L_a$ as in Fig.~2a, and this
$L_a$ is then a possible configuration for a section of $L$.

We still have to know in what order these different sections $L_a$
are joined together to form the tree $L\in\{T;\}$.
If we read back two paragraphs ago the description how $a[L]$ is folded to
get the derived amplitudes, it is clear that $K_1$ has to be the section
of $L$ containing the largest number $t_n$, and $K_2$ the section containing
the largest number of $L$ outside of $K_1$, etc. Given that, it is now
also clear that the tree $K=[K_1|K_2|\cdots |K_m]$ is a possible tree $L$.
Moreover, these $K_a$ are simply the partitions of $K$ stated just before
Theorem 5.

We can put together now all the ingredients for the proof of Theorem 5.
Start from the non-abelian amplitude $aV\{T;\}=\sum_{L\in\{T;\}}a[L]V[L]$
and use the unfolding theorem to expand $a[L]$. Next, group the result
by the derived amplitudes $\prod_aa[K_a]$, and sum over $L$. The coefficient
for each $a[K_a]$ is just
\begin{eqnarray}
\sum_{L_a\in{\cal U}[K_a]}\eta[L_a]V[L_a]=V^{MS}[K_a]\ .
\end{eqnarray}
That this sum is indeed the multiple commutator follows immediately
from the various definitions. Now putting all the different $a$ together
in the order discussed in the last paragraph, we finally arrive at
the right-hand side of (2.26), and hence Theorem 5.

\subsection{ Proof of Theorem 6}

The proof is identical to Theorem 5, except that $a$ must be replaced by
$a^*$, and the off-shell version of the folding theorem (Theorem 4) must
now be used.

\section{Acknowledgements}

This work was initiated during a visit of CSL to the University of Kentucky.
He would like to thank his colleagues there for hospitality. He also thanks
Y.J. Feng for drawing the figures in the paper. The research by CSL
is supported
in part by the Natural Science and Engineering Research Council of Canada,
and the Fonds pour la Formation de Chercheurs et l'Aide \`a la Recherche
of Qu\'ebec, and the research by KFL is supported by USDOE grant
DE-FG05-84ER40154.


\begin{thebibliography}{99}
\bibitem{GH} G. 't Hooft, {\it Nucl. Phys. B} {\bf 160} (1974) 461;
E. Witten, {\it Nucl. Phys. B} {\bf 160} (1979) 57;
S. Coleman, Erice lecture notes (1979).

\bibitem{JL} J.-L. Gervais and B. Sakita, {\it Phys. Rev. Lett.} {\bf 52}
(1984)
87; R. Dashen and A.V. Manohar, {\it Phys. Lett.} {\bf B315} (1993) 425, 438;
E. Jenkins, {\it Phys. Lett.} {\bf B315} (1993) 431, 441, 447;
R.F. Dashen, E. Jenkins and A.V. Manohar, {\it Phys. Rev. D}
{\bf 49} (1994) 4713.

\bibitem{CC} C. Carone, H. Georgi, and S. Osofsky, {\it Phys. Lett.} {\bf B332}
(1994)227;
M.A. Luty and J. March-Russell, {\it Nucl. Phys.}
{\bf B426} (1994) 71; M.A. Luty, {\it Phys. Rev. D} {\bf 51} (1995) 2322;
M.A. Luty, J. March-Russell and M. White, {\it Phys. Rev. D} {\bf 51} (1995)
2332; R.F. Dashen, E. Jenkins and A.V. Manohar, {\it Phys. Rev. D}
{\bf 51} (1995) 3697.

\bibitem{MP}
M.P. Mattis and M. Mukerjee, {\it Phys. Rev. Lett.} {\bf 61} (1988) 1344;
M.P. Mattis and E. Braaten, {\it Phys. Rev. D} {\bf 39} (1989) 2737;
P.B. Arnold and M.P. Mattis, {\it Phys. Rev. Lett.} {\bf 65} (1990) 831;
M.P. Mattis, {\it Phys. Rev. D} {\bf 51} (1995) 3267.

\bibitem{YG} Y.G. Liang, B.A. Li, K.F. Liu, and R.K. Su, {\it Phys. Lett.}
{\bf B243} (1990) 133.

\bibitem{CS} C.S. Lam, `Soft interaction of heavy mesons', Proceedings
of the Heavy Quark Workshop, Institute of Theoretical Physics, Beijing,
August, 1995; C.S. Lam and K.F. Liu, to be published.

\bibitem{YJ} Y.J. Feng, O. Hamidi-Ravari, and C.S. Lam,
`Cut diagrams for high energy scatterings', to be published.

\bibitem{HC} H. Cheng and T.T. Wu, {\it Phys. Rev. Lett.} {\bf 22} (1969)
666; {\it Phys. Rev.} {\bf 182} (1969) 1868, 1899; M. Levy and J.
Sucher, {\it Phys. Rev.} {\bf 186} (1969) 1656.

\bibitem{DR} D.R. Yennie, S.C. Frautschi, and H. Surra, {\it Annals of Phys.}
{\bf 13} (1961) 379; G. Grammer and D. Yennie, {\it Phys. Rev.} {\bf D8}
(1973) 4332.

\bibitem{HCT} H. Cheng and T.T. Wu, Expanding Protons: Scattering at High
Energies (MIT Press, 1987).

\bibitem{SW} S. Weinberg, {\it Phys. Rev. Lett.} {\bf 16} 879 (1966);
{\it Phys. Rev.} {\bf 177} (1969) 2604;
{\it Phys. Rev. D} {\bf 2} (1970) 674, 3085;
{\it Phys. Rev. Lett.} {\bf 65} (1990) 1177;
L. Brown, {\it Phys. Rev. D} {\bf 2} (1970) 3083.
\end{thebibliography}
\end{document}